# The Emergence of Chirality from Metabolism

John F. Malloy[1], Camerian Millsaps[1], Kamesh Narasimhan[2], Louie Slocombe[2], Cole Mathis[2,3,4], Leroy Cronin[5,6], Sara Imari Walker[1,2,6]*

[1]School of Earth and Space Exploration, Arizona State University, Tempe AZ USA

[2]Beyond Center for Fundamental Concepts in Science, Arizona State University, Tempe AZ USA

[3]School of Complex Adaptive Systems, Arizona State University, Tempe AZ USA

[4]Biodesign Institute, Arizona State University, Tempe AZ USA

[5]School of Chemistry, University of Glasgow, Glasgow, G12 8QQ, UK

[6]Santa Fe Institute, Santa Fe, NM USA

*author for correspondence: sara.i.walker@asu.edu

# Abstract

Molecular chirality is critical to biochemical function, but it is unknown when chiral selectivity first became important in the evolutionary transition from geochemistry to biochemistry during the emergence of life. Here, we identify key transitions in the selection of chiral molecules in metabolic evolution, showing how achiral molecules (lacking chiral centers) may have given rise to specific and abundant chiral molecules in the elaboration of metabolic networks from geochemically available precursor molecules. Simulated expansions of biosphere-scale metabolism suggest new hypotheses about the evolution of chiral molecules within biochemistry, including a prominent role for both achiral and chiral compounds as nucleation sites of early metabolic network growth, an increasing enrichment of molecules with more chiral centers as these networks expand, and conservation of broken chiral symmetries along reaction pathways as a general organizing principle. We also find an unexpected enrichment in large, non-polymeric achiral molecules. Leveraging metabolic data of 40,023 genomes and metagenomes, we analyzed the statistics of chiral and achiral molecules in the large-scale organization of metabolism, revealing a chiral-enriched "phase" of network organization evidenced by system-size dependent chiral scaling laws that differ for individuals and ecosystems. By uncovering how metabolic networks could lead to chiral selection, our findings open new avenues for bridging metabolism and genetics-first approaches to the origin of chirality, allowing tools for better timing of major transitions in molecular organization during the emergence of life, understanding the role of chirality in extant and synthetic metabolisms, and informing targets for chirality-based biosignatures.

## Introduction

Identifying the sequence of evolutionary events that gave rise to the last universal common ancestor (LUCA) of cellular life[1] will be crucial to elucidating the origin of life on Earth. However, the timing of many important major transitions in molecular structure bridging geochemistry to LUCA remain uncharacterized[2]. Historically, the dominant sets of hypotheses about these earliest steps fall under either metabolism-first[3,4] or genetics-first[5,6] scenarios, depending on whether the hypothesis favors selection of reproducing small molecule network organization (metabolism), or replicating polymers (genetics) as the first of the significant early molecular transitions. However, there have recently been calls for more integrated approaches, which do not fall neatly within these traditional divisions[7,8]. Genetics and metabolism, as understood in modern cells, are potentially late innovations in the evolution from geochemistry to LUCA[9] and are, in any case, *biologically derived* categories[10] defined by LUCA's cellular architecture[11]. These distinctions may not strictly apply to the systems that preceded it. More basic questions remain open about what *chemically derived* properties might have driven the emergence of LUCA via evolutionary processes that predate the genome.

One chemical property that transcends specific hypotheses about the detailed chemistry of life's origin is molecular chirality[12]. Among the major transitions in molecular structure leading to LUCA, one of the most distinctive and widely studied is the onset of macromolecular homochirality[13]. Proteins are made of left-handed (L) amino acids, while RNA and DNA are composed of right-handed (D) sugars[14]. Homochirality has long been regarded as essential to biological functionality[15], though some studies suggest that small proteins composed of mixed chirality amino acids could also be functional[16]. Importantly, abiotic systems, including meteoritic samples[17,18], do not display the strong degree of chiral selection observed in biology. Several viable mechanisms[19–25] have been proposed to break symmetry between mirror-image (L or D) enantiomers. Most of these mechanisms are studied in the context of prebiotic experiments probing genetics-first hypotheses[26–29], although some studies have explored bridges between the chirality of nucleotides and amino acids and potential metabolites that could support their asymmetric synthesis[30–33]. However, one open question is if a chiral excess is required for the origin of life, or if instead the origin life event was associated with the generation of chiral molecules and the chiral phase transition(s) leading to the situation we observe today[34].

Chirality plays such a pervasive role in Earth's chemistry and biochemistry[12], both naturally and synthetically, it is easy to assume chiral centers (positions in a molecule where a mirror-symmetry is centered) are a common feature of synthesizable molecules on all planetary bodies. However, many of the most abundant geochemical produced species, like $CH_4$ and $CO_2$, are achiral[35], and in geochemistry, it is generally not expected that chiral species will be nearly as enriched as what is found in biology, nor as enriched as what is included as purified reagents for prebiotic chemistry experiments[36]. Unconstrained prebiotic mixtures, such as variations of the classic Miller-Urey experiment[37] produce both achiral and chiral molecules, where the product achiral species are abundant, and the chiral molecules are predominantly present in mirror-image forms as racemic, or very near racemic, mixtures[38]. This is similar to what is observed in extraterrestrial samples[17,18,39]. Potential metabolic precursors to life contain few, if any chiral centers: within the reductive citric acid cycle (rTCA), hypothesized as among the earliest metabolisms and potentially predating LUCA[2], only malate and isocitrate are chiral. Getting from abundant small, achiral molecules intermixed with lower abundances of slightly heavier, racemic chiral molecules, to the very large, abundant macromolecular homochiral polymers of LUCA leaves a wide gap with many open questions.

A striking pattern[40,41] emerges when one considers the structure of chemical space: the low molecular weight compounds we might expect to form most readily from planetary environments in the absence of life, are dominantly achiral, and contain no chiral centers. Recent models have shown that large chemical systems, with many chemical species of high molecular weight (many heavy atoms), might be necessary to drive spontaneous symmetry breaking to a homochiral state[41]. Most known mechanisms[19,22,24,25] require as a prerequisite that chiral molecules are present, and in high abundance. Taken together with what is known of abiotic chemistry, these insights suggest a missing stage in chemical evolution: before the onset of homochirality, there may have been another major transition, from a predominantly *achiral*, abiotic chemical universe to one with abundant enough chiral species for spontaneous symmetry breaking between mirror image enantiomers to even occur[41]. This opens new hypotheses that could link metabolic and genetic approaches to the origin of life, and allow addressing the question: what came first, chirality or life?[42]. The vastness of chemical space suggests that reaction pathways that produce abundant and specific chiral molecules may be a prerequisite to subsequent transitions in the evolution of geochemistry into biochemistry, including the origin of homochirality by spontaneous symmetry breaking.

Our goal in the current work is to shed light on missing stage(s) of chiral molecular evolution, by bridging a predominantly achiral abiotic world with limited chiral selection, to the (racemic) chiral dominated world, which broke symmetry to yield the homochiral macromolecular systems characterizing LUCA. We view modern metabolism as holding clues about this bridge. To this end, we adopt a network and statistics-based approach to study the progression and distribution of chiral molecules within extant metabolic networks, providing evidence for the evolution of chiral molecules in metabolism, and new biosignatures in the network-scale organization of chiral molecules.

## Results

Chiral centers are a specific subset of stereocenters[43] where four distinct substituents are bound to the central atom. Here, we explore the plausibility of an achiral-to-chiral transition, placing early metabolic evolution within the broader context of chemical space, **Figure 1A**. For the purposes of this work, we use the presence or absence of chiral centers to classify molecules as chiral or achiral. Molecules with no identifiable chiral centers are categorized as ***achiral***, as they lack the structural features necessary for chirality, whereas molecules with at least on chiral center are categorized as ***chiral***. We define a ***chiral order parameter,*** $\mathbb{C} = \frac{C}{A+C}$, to quantify the fraction of chiral compounds in a sample, where $C$ and $A$ correspond to the numbers of chiral and achiral molecular species, respectively. In the theory of phase transitions, order parameters arise from symmetry-breaking where extra variables become necessary to describe the state of the system, and these typically range from a value of 0 in one phase to non-zero in another[44]. The broken symmetry explored here occurs in the formation of chiral centers, where more than one non-superimposable 3D conformation of a given chiral center becomes possible for the same molecular formula[12].

Observation of an abundant chiral molecule, therefore, represents selectivity in the reaction pathway that produced it, where a specific 3D conformation is synthesized at an entropic cost related to the number of stereoisomers that correspond to the same molecular formula. The two limiting regimes are ensembles of dominantly achiral molecules (with no chiral centers), when $\mathbb{C} \to 0$, or dominantly chiral molecules (where all molecules have at least one chiral center), when $\mathbb{C} \to 1^{+/-}$. We use the notation +/– here to denote that the chiral phase could correspond to a racemic system with both enantiomers present (+ –) or a homochiral system that is all L (+) or all D (–) enantiomers. For example, prebiotic chemistry experiments starting from racemic chiral mixtures would have as a starting condition $\mathbb{C} = 1^{+-}$, and end in a homochiral-L ($\mathbb{C} = 1^{+}$) or homochiral-D ($\mathbb{C} = 1^{-}$) at the onset of homochirality. For brevity of notation, we use $\mathbb{C} = 1$ in most of what follows, as our data annotation identifies chiral centers but not their specific enantiomeric connotation.

The size of chemical space grows *at least* exponentially with the number of heavy atoms in a molecule[45], meaning it is not possible to elucidate all possible small chemical structures computationally, nor in physical reality with the resources on a planet[46]. We, therefore, sampled an enumeration of the Generated Database (GDB)[45] to study chemical space for molecules with ≤18 heavy atoms, **Figure 1B,** which samples a well-defined space of molecules without constraint of whether these haven been synthesized. $\mathbb{C} \sim 0$ for molecules with very few heavy atoms (see SI Section 1.4 for details on enumeration and sampling), and there is an abrupt phase boundary, separating an achiral and chiral phase, located at about 8-9 heavy atoms. Below the boundary, nearly all molecules are achiral, and above it, nearly all molecules have one or more chiral centers, with $\mathbb{C} \to 1$ for molecules with ~15 heavy atoms or more (SI Sections 1.3 & 1.4), confirming earlier findings[40,41]. By attaching only a few atoms via new bonds, novel 3D symmetries are introduced[43], and nearly every achiral molecule just below the phase boundary can be transformed into a chiral molecule above it. This generic feature of chemical space is highly relevant in a messy prebiotic context[47]. If enough achiral molecules near the phase boundary, containing ~8 heavy atoms, can persist in a geochemical environment, nearly all new bonds formed, even with unconstrained reactions[47], would be statistically likely to produce chiral centers.

Biochemistry occupies a very small region of chemical space. There are an estimated $10^{60}$ possible molecules composed of C, O, N, and S made of up to 30 atoms[46]. By contrast, there are only 18,716 cataloged and verified small molecules in the Kyoto Encyclopedia of Genes and Genomes (KEGG)[48] database (as of writing), which catalogs known metabolites and biochemical reactions. These reactions are catalyzed by proteins, composed of homochiral molecular subunits, which critically depend on these reactions for their own synthesis. Thus, one hypothesis is that biochemical space should be predominately chiral, occupying parts of chemical space where $\mathbb{C}\sim 1$, because the modern functioning of biochemical systems crucially relies on chiral molecules, and their reliable synthesis, for protein-mediated reactions to occur at all. However, studies of how biochemistry emerged from geochemistry suggest some subsets of metabolic networks may be very ancient[4,49,50], dating back to the early evolutionary steps in the origin of life, and these in general contain few chiral molecules[2]. Therefore, an alternative hypothesis would predict metabolism to retain a similar transition regime, reflecting its selection from the larger chemical space available on planets[51], which should itself exhibit evidence of a phase boundary between achiral and chiral-dominated chemical space.

We thus sought to determine if similar evidence of an achiral-to-chiral transition regime is evident within the much smaller set of chemical space occupied by metabolism (SI Section 1.2). We find a transition in metabolic space in a similar heavy atom count range as in estimates for broader chemical space, **Figure 1C,** but interestingly the phase transition in metabolic space is frustrated relative to what is found in chemical space more broadly, exhibiting more heavy achiral molecules than statistically expected based on naïve chemical space estimates, **Figure 1D** (see also SI Figure 2). We use the term "frustrated" in the sense of physical phase transitions, where phase transitions in physical materials often do not run to their theoretical completion due to interaction and finite system effects. In the chiral phase transition, we anticipate that interaction effects arise from reaction chemistry and network organization, constructed through the process of evolution. Selectivity can be low in the achiral phase; however, for chemical systems to persist in the chiral phase, they must exhibit a high degree of selectivity against the combinatorial explosion of possible molecular structures. There is a higher entropic cost associated with the synthesis of a specific 3D structure when, as the number of heavy atoms per molecule increases, there is a corresponding and rapidly growing number of structural forms with the same heavy atom composition. That is, the frustrated effect in real systems may be driven by molecules being part of functional reaction networks, such as those found in biology, where precision in 3D structure is crucial for function. The surprisingly enriched, large, non-oligomeric achiral molecules in metabolic space, which are rare in chemical space, are also potentially indicative of a signature of biochemical selectivity, thus informing new chirality-based biosignatures.

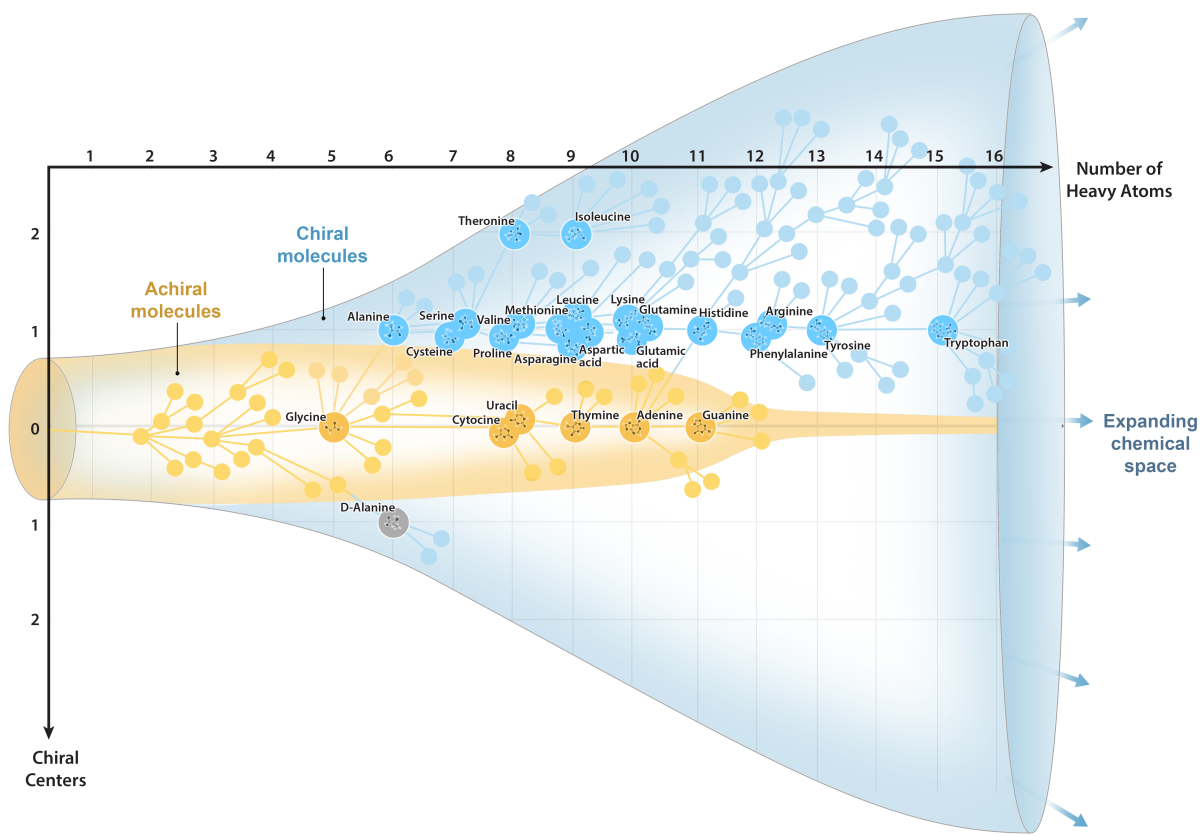

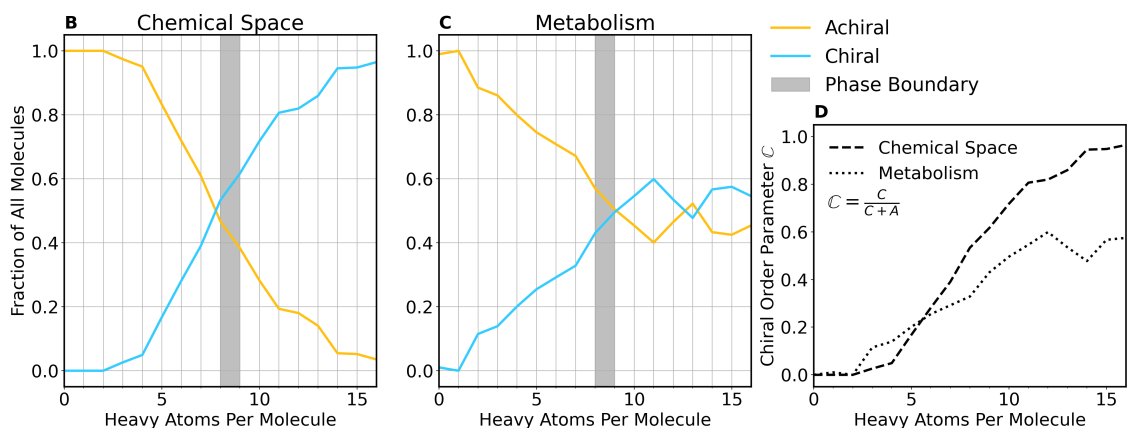

**Figure 1:** A molecular-size dependent transition in the likelihood of achiral ($A$) and chiral ($C$) molecules within chemical space. **(A)** As heavy atoms are added to molecular graphs in an expanding chemical space, the total number of molecular configurations grows at least exponentially, and an abrupt transition marks where chiral centers become statistically more likely to be present[40,41]. **(B)** Structural elucidation of molecules composed of up to 16 heavy atoms, including C, N, O, S, and halogens from the Generated Database (GDB)[45], reveals a transition point at ~8 heavy atoms, where the space of possible chemical

structures yields vastly more chiral compounds than achiral ones. By ~15 heavy atoms, nearly all molecules are chiral, and the ratio of chiral to achiral compounds, $A/C \rightarrow 1$. **(C)** Considering only the biologically selected chemical space found in metabolisms, as cataloged in the Kyoto Encylcopedia of Genes and Genomes (KEGG)[48] database, a similar transition is found at ~9 heavy atoms, where chiral molecules become more prevalent. For larger size metabolites, achiral molecules are much more prevalent then naïve chemical space estimates would predict, suggestive of new biosignatures based on identification of statistics of chiral and achiral molecular graphs. **(D)** Comparison of GDB and KEGG data shows how metabolism is enriched relative to chemical space in heavy achiral molecules, lowering the chiral order parameter, $\mathbb{C}$, observed in heavier molecules, and providing potential evidence for larger non-oligomeric achiral molecules as a biosignature.

## Chirality in the Expansion of Metabolism

To explore the potential for an achiral-to-chiral phase transition in a network and evolutionary context, we modeled the progression of geochemistry into biochemistry by tracking an expanding network of cataloged metabolic reactions starting from geochemically plausible precursor molecules[52,53]. We implemented a network expansion algorithm (SI Section 2.2) on a bipartite, undirected network of planetary-scale metabolism (SI Section 2.1). Similar large-scale network constructions of metabolism have been used to study biochemical network organization and its change over geological timescales, providing insights into the metabolic evolution of organisms and the entire biosphere[49,54], the viability of metabolisms across diverse ecological and planetary environments[55–57], and testing prebiotic hypotheses regarding the earliest evolution of metabolism[58–60]. To investigate the evolution of chirality via network expansion, we used a newly developed, carefully curated database, CBR-db[61], which is derived from KEGG[48] and the ATLAS of Biochemistry[61,62]. CBR-db offers an advantage over other databases for our study because all reactions are carefully curated to be stoichiometrically balanced and include additional chemical and physical requirements for viable reaction chemistry. To be as comprehensive as possible, covering the wide range of more detailed and specific hypotheses about the geochemistry that preceded life, we also use a larger set of geochemically plausible compounds than in prior work on network expansion models of the transition from geochemistry to biochemistry[59,63] (see SI Section 2.3). We aim to investigate the mechanisms at play that might be universal to any origin of life event, rather than those specific to a given geochemical context or emergently selected molecular structures.

Network expansion[52,53] allows exploring percolation through the space of characterized metabolic reactions by following successive addition of reactions to a growing network. We note that because we are performing expansion on the modern network, we are not aiming to recapitulate exact steps of early evolution but instead are looking for general principles in the statistics of biochemical network organization, which could inform origin of life models and experiments. In network expansion, each iteration of the algorithm constitutes a "generation" that adds new reactions (SI Section 2.2). We begin with a set of 62 achiral compounds that are plausible to have persisted in the prebiotic environment (SI Section 2.3 and SI Table S1), referred to herein as the "primordial seed set" to explore how metabolic reactions can expand from abiotic, achiral molecules. We tracked the chiral order parameter $\mathbb{C}$, and the cumulative number of chiral and achiral compounds $C$ and $A$ through the network expansion process, as shown in **Figure 2A**, finding that between generation 3 and 4 of the expansion process, the global metabolic network transitions from being predominantly achiral to dominantly chiral (**Figure 2B & C**). Chiral molecules first appear in generation 1 of the primordial expansion algorithm. The transition point $\mathbb{C} \sim 0.5$ (representing roughly equal numbers of chiral and achiral species) occurs where the average heavy atom count per molecule is 7.60 amu, as predicted by chemical space estimates[41], **Figure 2C**. We annotated reactions using a metabolic model of LUCA (**Figure 2B, purple bars**) derived from a consensus assessment of eight proposed genomic models developed previously[64]. Given that LUCA was homochiral, this allows us to estimate a network subgraph where the homochiral molecular transition must have already completed, which can be labeled in the expanding network. When the first chiral molecule appears in generation 1 of the expansion, there are 221 compounds and 1973 reactions present. In generation 18, when all LUCA compounds accessible to the expansion are present (SI Section 2.6), there are 8,481 compounds and 113,618 reactions. Over this span of 17 expansion generations, we observe a transition from achiral-dominated ($\mathbb{C} \sim 0$), to chiral-dominated ($\mathbb{C} \to 0.68$).

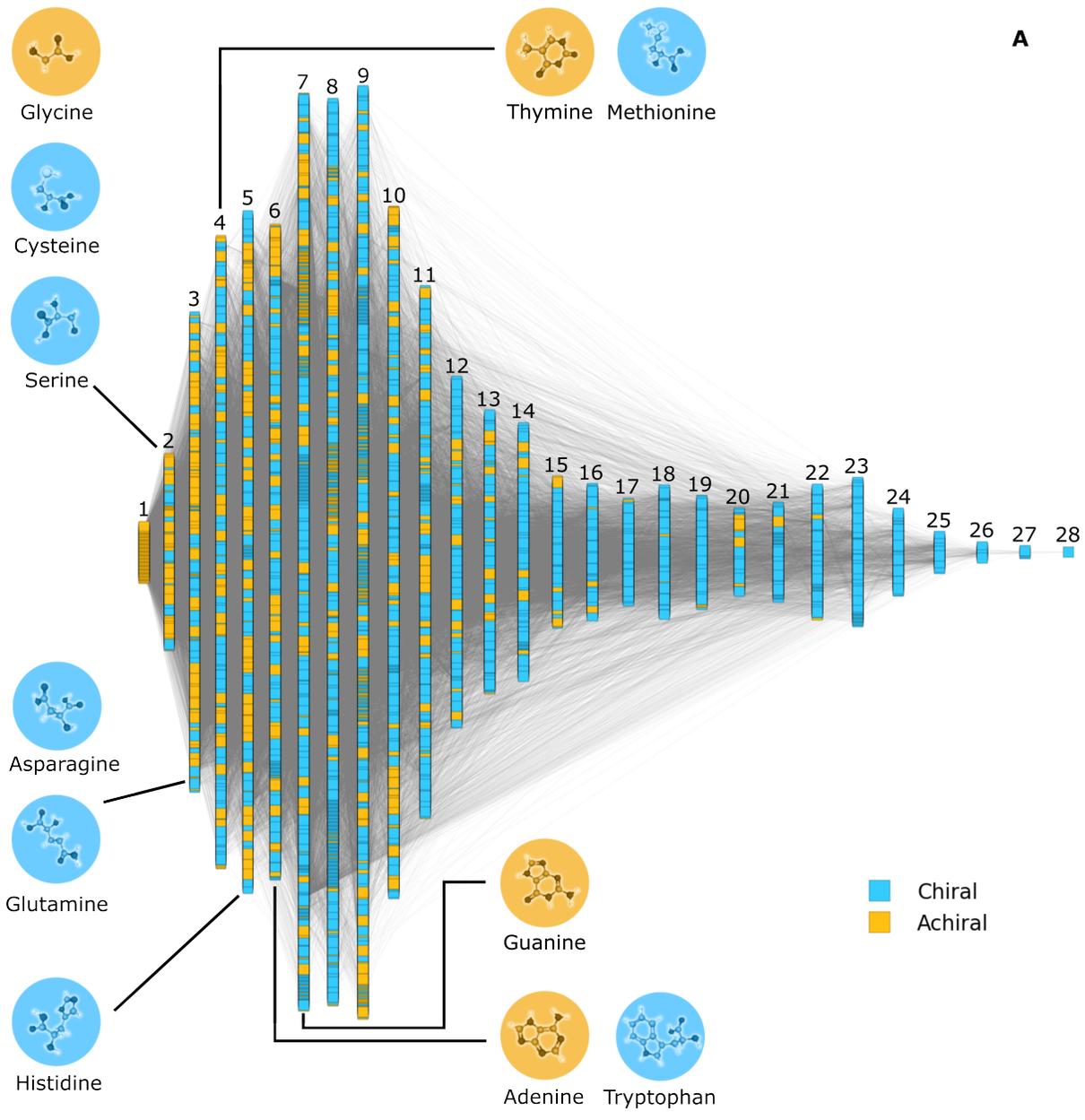

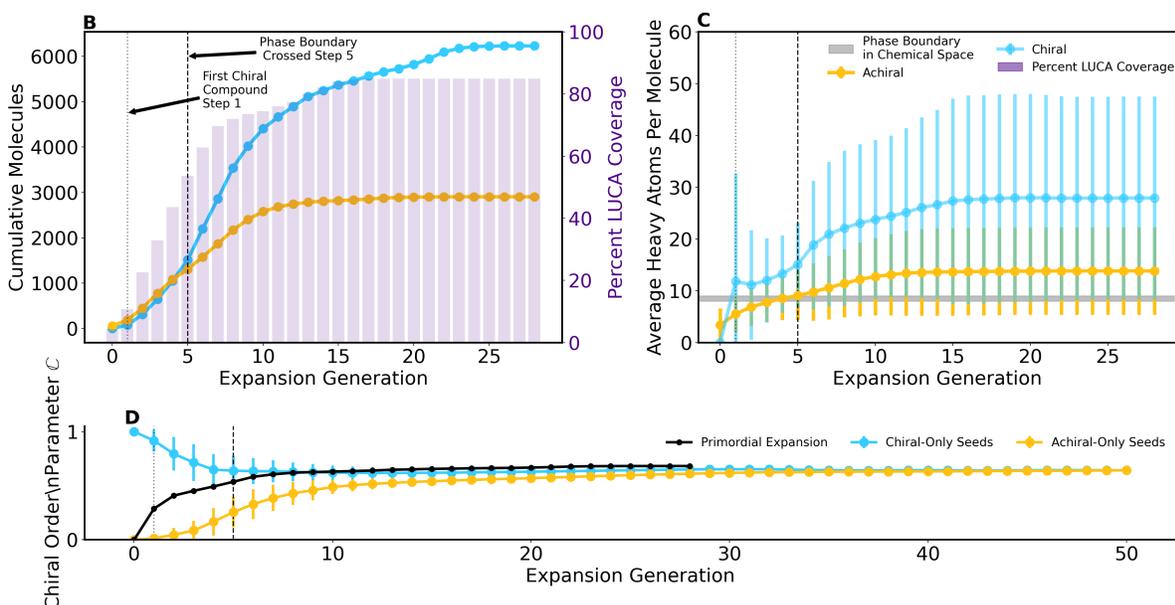

**Figure 2:** Network expansion reveals evidence for an achiral (orange) to chiral (blue) transition in the evolution of metabolism. **A)** Visualization of the expanding network, where each generation is represented by the number of new molecules added that generation (colored as chiral or achiral), and edges connect reactants to products in subsequent generations. **B)** Expanding from a primordial seed set of achiral, prebiotically plausible starting molecules, we track the cumulation of chiral and achiral molecules during the network expansion, including the percentage coverage for a consensus model of LUCA metabolism (purple). A steep increase in chiral molecules begins at generation 3, resulting in an inflection point where the number of chiral molecules surpasses that of achiral ones, starting at generation 4. **C)** Average and standard deviation of the average number of heavy atoms in achiral and chiral compounds generated during each generation of the primordial expansion shows the mean size of achiral compounds at the transition point $\mathbb{C} \sim 0.5$ to be ~7.6 amu, consistent with the phase transition range in chemical space (Figures 1, B, and C), with chiral molecules typically larger. **D)** Starting from randomly sampled all-chiral or all-achiral seed sets reveals how chiral seed sets converge more quickly to the biosphere's chiral order parameter, indicative of an important role for chiral molecules as central to the expansion process.

Network expansion is known to be sensitive to seed set[52]. We therefore analyzed the expansion process for randomly sampled seed sets containing exclusively chiral or exclusively achiral molecular species, with 100 seed sets sampled for each set size between 2 and 10 molecules, yielding a total of 1,800 randomly initialized expansions (see **Figure 2D,** and additionally SI Section 2.3 for more details**)**. Our results are robust to seed set choice, indicating they are a generic structural feature of metabolism. We find that when starting from exclusively chiral compounds, expansions converge to the trajectory of the primordial expansion much more quickly than what is observed for randomly seeded expansions starting with achiral seeds. The likelihood of convergence to the full network, and the number of generations necessary, depends on seed set size, and on average we find randomly generated chiral seed sets can be smaller (~20 molecules) than achiral ones (~50 molecules) and still converge to the full network (see SI Figure 5). This could be because chiral molecules tend to be heavier and more complex, whereas most achiral seed sets generated by random sampling will lack heavier achiral compounds near the phase boundary (e.g., > 7 heavy atoms) where their likelihood to produce chiral molecules increases. To test the role of heavy achiral molecules, we conducted another set of expansions, starting from the primordial expansion set modified by selectively dropping achiral species with a heavy atom count greater than a given threshold (SI Section 2.4). Smaller achiral molecules like $H_2O$, $O_2$ and $H^+$ from the primordial seed set are dominant hubs in the network, so we might expect the presence of these and other small, prebiotically plausible achiral molecules to be sufficient to drive the network to its full scope. However, we find that including only those achiral compounds with < 4 heavy atoms in the primordial seed set (35 molecules) leads to final network sizes that are significantly smaller the final networks generated from expansions which include heavier achiral molecules (see SI Figures S8 & S9). This indicates larger achiral molecules (>4 heavy atoms) play a role in accessing the full chemical space of Earth's modern metabolism, supporting our hypothesis that larger achiral compounds - near the achiral/chiral phase boundary - may increase the statistical likelihood of making chiral molecules more readily in expanding metabolic networks.

Taken together, the network expansions provide one possible explanation for the frustrated achiral-to-chiral phase transition observed in **Figure 1C**. Unlike random samples of chemical space, evolving reaction networks must grow contingently by constructing new molecules from extant ones via the expansion of, and selection on, existing reaction pathways. Since the simplest molecules are often achiral, this suggests that slightly heavier molecules acted as nucleation sites, driving the very earliest stages of evolution through achiral molecules closely connected to chiral ones. Some of these molecules later became hubs in the expansion of metabolism as selectivity increased along specific synthesis pathways. Chiral selectivity could have emerged as reaction pathways connected to chiral hubs broke symmetry when networks began to solidify with catalytic specificity against a background of unconstrained and messy geochemical reactions.

To confirm the critical role for chirality in nucleation sites for early metabolic network expansion, we studied the topological structure of the network, with molecular nodes labeled as chiral or achiral. Visualizing the network, **Figure 3A**, shows highly connected achiral hubs, which were present in the primordial seed set. But many hubs are also chiral. These chiral hubs could have played a prominent role as nucleation points through which life emerged as metabolism expanded into larger, more complex regions of chemical space defined by more heavy atoms per molecule. We labeled the number of chiral centers per molecule, which is 0 for achiral molecules and $\geq 1$ for chiral molecules (with one or more chiral centers). We used this labeling to track molecules based on their "chiral complexity", quantified by the number of chiral centers per molecule. We confirm many of the highest-degree nodes (SI Table 2) are chiral. Some important high-degree nodes include multiple chiral centers, **Figure 3B**, like nicotoinamide adenine dinucleotide (NAD) and related compounds (NADH, NAD+, NADP+, NADPH) and ATP and AMP. This is particularly significant given how, overall, node degree tends to decrease with the number of chiral centers. Forming chiral molecules requires specificity that increases with the number of chiral centers per molecule. Naïvely, we might therefore expect that molecules with high chiral complexity should be more likely to appear later in metabolic evolution[65–67], e.g., once chiral selective catalysts evolved which are specific enough to counter the entropic cost of chiral specificity[68]. The more general trend outside of these hubs is that achiral molecules or those with few chiral centers are predominantly added early, and the upper bound on the number of chiral centers in any molecular species increases throughout the expansion process (**Figure 3C**). This trend confirms how metabolic networks construct increasingly chiral specific synthesis pathways and, therefore, chiral complexity as the encoded network expands.

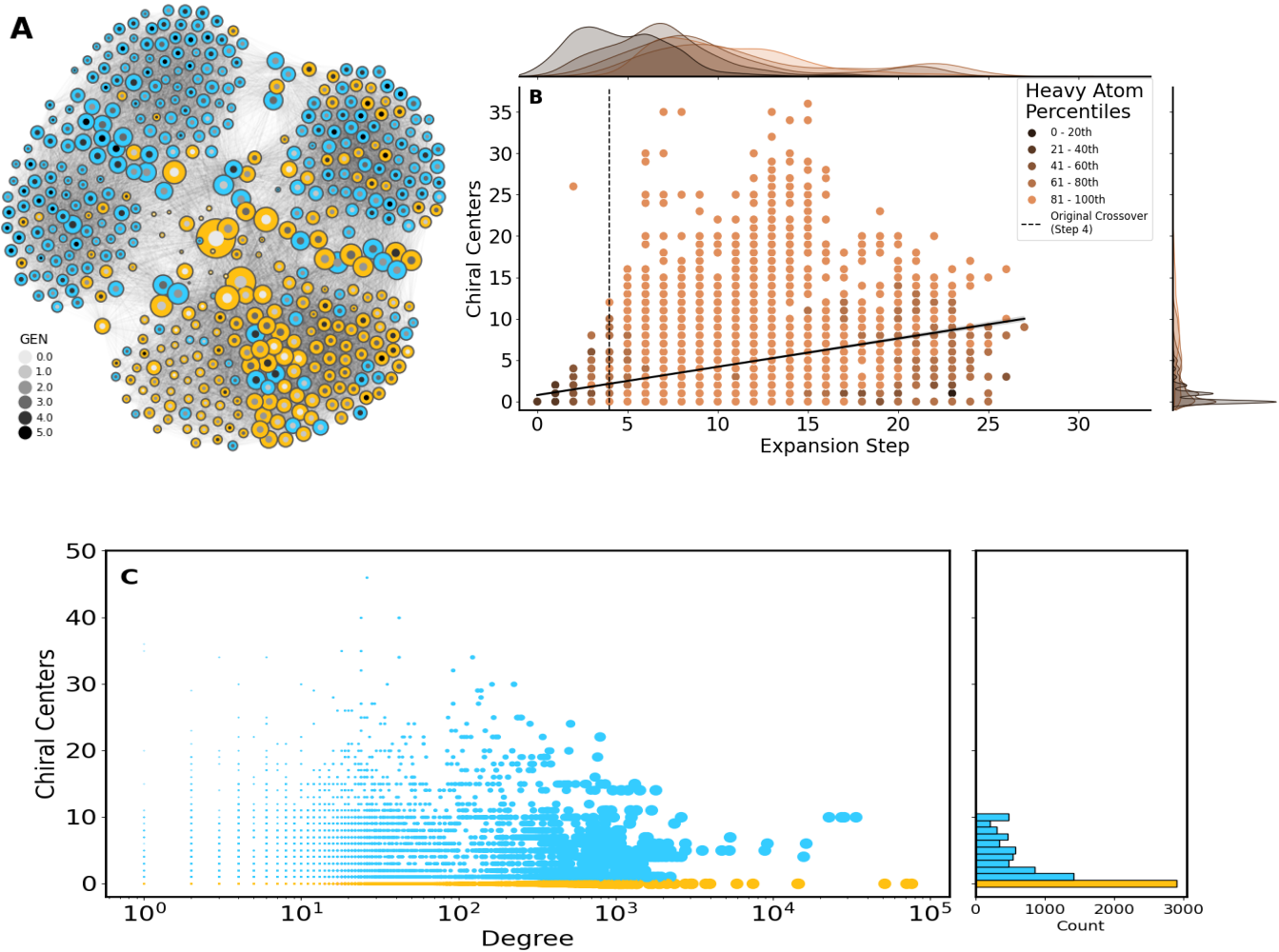

**Figure 3:** Chiral hubs play a prominent role in metabolic networks, with the potential for larger molecules with more chiral centers to be produced as metabolism expanded. **A)** Snapshot of the metabolic network at generation 5 of the primordial expansion, where the phase boundary at $\mathbb{C} = 0.5$ has been crossed, and the network first becomes dominated by chiral molecular species. Molecular node sizes are scaled to reflect relative edge counts (number of reactions), and the network is pruned to remove nodes with fewer than 50 edges for tractability in its visualization. Edges connect reactants with products. Achiral compounds like $H_2O$ and $CO_2$, found in the primordial seed set, form prominent hubs, but many chiral hubs are also evident. **B)** Tracking chiral centers per molecule within the expanding network over time reveals a rate increase of 0.34 chiral centers added to the highest chiral complexity molecules each generation. Large molecules, with many heavy atoms (measured here in percentiles of atomic weights), appear more frequently in later expansion steps. **C)** The degree distribution shows the highest degree molecules in the network are achiral species (the three highest degree nodes are $H_2O$, $H^+$, and $O_2$, respectively (SI Table S2)) but other prominent hubs (NAD, NADH, NAD+, NADP+, NADPH, AMP, ATP) have several chiral centers. Over the entire network, degree tends to decrease with the number of chiral centers, corroborating expansion results that indicate chiral complexity increases as the network expands.

## Regime Shifts in Chiral Scaling Laws

It is a common feature in the theory of physical phase transitions that different phases are associated with different scaling regimes for system properties[44]. In biology, identification of shifts in scaling behavior have seen widespread application to the study of evolutionary transitions[69]. As such, scaling relationships associated with the use of chiral compounds across metabolisms might provide clues to the evolutionary transitions in molecular networks that played a role in the emergence and evolution of life. We therefore calculated $\mathbb{C}$ across a sample of 40,023 genomes and metagenomes from the Joint Genome Institute's Integrated Microbial Genes and Genomes database (JGI IMG/M)[70], differentiated as individual taxa, labeled by domain as Archaea, Bacteria or Eukarya, or metagenomes, corresponding to metabolic networks functioning at the scale of individual organisms or entire ecosystems, respectively. We plotted the system-size dependency of $\mathbb{C}$ for these data, where network size is determined by the number of unique molecules encoded in the metabolic network of the genome or metagenome (SI Section 3.1). We find a consistent scaling pattern across genomes sampled from Bacteria, Archaea, and Eukarya, where the chiral order parameter $\mathbb{C}$ systematically *decreases* from $\mathbb{C} \sim 0.75$ for the smallest networks to $\mathbb{C} \sim 0.55$ for the largest (**Figure 4A**). Conversely, for metagenomes, the scaling trend is opposite, with a systematic *increase* in the chiral order parameter from $\mathbb{C} \sim 0.55$ for the smallest samples to $\mathbb{C} \sim 0.62$ for the largest. We also computed $\mathbb{C}$ for the consensus metabolism of LUCA[64] to be $\mathbb{C} = 0.65$, and for the biosphere (KEGG) to be $\mathbb{C} = 0.63$. While $\mathbb{C} = 0.5$ clearly marks the phase boundary in chemical space, the bulk $\mathbb{C}$ of small, synthesizable molecules, averaged across different size molecules could be quite different because of the constraints imposed on physical molecules by plausible reaction chemistry of Earth and in the lab. To have a standard average value of *synthesizable chemical space* for more direct comparison, we performed elemental-based sampling of cataloged molecules in Reaxys[71], one of the world's largest databases cataloging known molecules (SI Section 4.2). We find a lower bound of $\mathbb{C} \sim 0.54$ from randomly sampling molecules in Reaxys with the restriction the molecules can only use elements found in biological metabolism, and an upper bound of $\mathbb{C} \sim 0.57$ sampling only molecules made of CHNOPS elements (SI Section 4.2). That is, we find the bulk $\mathbb{C}$ sampling from all known molecules to sit just above the phase boundary. We also find most metabolisms sit at even larger values of $\mathbb{C}$, **Figure 4B**. Although the chiral scaling laws for metabolisms constructed from genomes and metagenomes display opposing trends, these converge at an inflection point, marking a scaling-regime shift within the range of $0.54 < \mathbb{C} < 0.57$. This chiral scaling transition between individual and ecosystem-scale metabolic networks intersects precisely within the range of bulk properties of the chemical space of all known molecules sampled more broadly. These results suggest a major shift in the molecular organization of metabolisms akin to major shifts in scaling observed in other biological transitions[69].

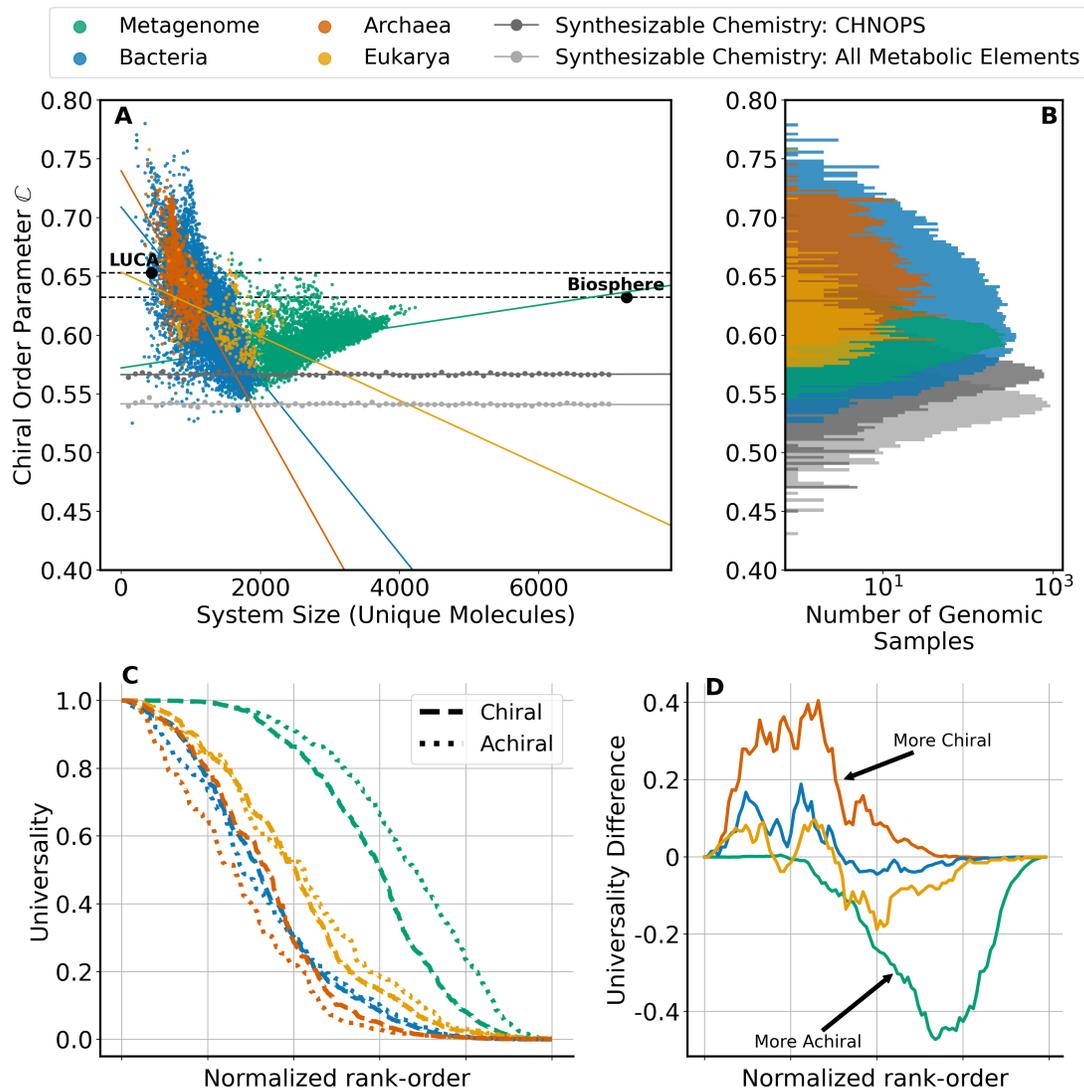

| Domain | Chiral AUC | Achiral AUC |
|---|---|---|
| Metagenome | 0.488 | 0.542 |
| Eukarya | 0.333 | 0.352 |
| Bacteria | 0.251 | 0.243 |
| Archaea | 0.214 | 0.166 |

**Figure 4.** A transition in scaling behavior for the chiral order parameter, $\mathbb{C}$, across metabolic networks from genome and metagenome data (the latter representing ecosystem-scale metabolisms, which are sampled from diverse environments). **A)** Systematic trends in individual taxa sampled from Archaea (red), Bacteria (blue), and Eukarya (orange) reveal scaling behaviors associated to a *decrease* in the chiral order parameter with increasing system size. The opposite trend is observed at the ecosystem-scale, where ecosystem-scale metabolic networks (green) exhibit a trend of *increasing* $\mathbb{C}$ with system size (see SI Table S4 for scaling coefficients). Data from these two opposing scaling regimes converges within a range consistent with bulk samples of the chemical space of known molecules, sampled from the Reaxys database (grey), where sampling was done restricted to molecules using the same elements as in metabolism, or CHNOPS elements only. Also included are data for a consensus metabolic model of LUCA with $\mathbb{C} = 0.65$, and the modern biosphere with $\mathbb{C} = 0.632$ (black points). **B)** Metabolic networks tend to be enriched in chiral molecules (larger $\mathbb{C}$) relative to bulk estimates of cataloged molecules, with genomic-scale metabolisms tending to be more chiral enriched than metabolisms sampled at the ecosystem-scale. **C)** Universality of chiral and achiral molecules across datasets. Area under the curve ($AUC$) scores are used to quantify universality, where $AUC \to 1$ indicates more universal molecules, and $AUC \to 0$ indicates more molecular uniqueness across samples. **D)** Comparing the universality of chiral and achiral molecules reveals how ecosystem-scale networks exhibit an enrichment of achiral molecules that are *less* universally distributed. The general trend, moving from the metabolisms of ecosystems to those of Eukarya, Bacteria, and Archaea, corresponds to one of decreasing network size and a corresponding increase in chiral enrichment of more universally distributed molecules. **E)** Table of AUC scores corresponding to panel C.

It is well documented that many molecules are universally shared across organisms, allowing for inference of the properties of LUCA. We aimed to determine whether universal achiral or chiral molecules might drive the different chiral scaling regimes. We constructed a rank-ordered plot of the frequency of molecules across samples and computed the area under the curve (AUC) for these distributions to compare the universality of chiral and achiral molecules (see SI Section 3). An AUC of ~1 indicates complete universality, meaning all molecules are present in every sample in the dataset. On the other hand, AUC = 0 indicates no universality, meaning all molecules are unique to a single sample. We find that metagenomic metabolisms rank higher in universality for achiral compounds. Eukarya metabolisms are approximately equivalent in the universality of achiral and chiral molecules, whereas chiral molecules are more universal in Archaeal and Bacterial metabolisms (**Figure 4C**). Achiral and chiral molecules are equivalently universal across metagenomic metabolisms for the most widely distributed molecules, but for less universally distributed molecules, it is the achiral molecules that are found across more samples in ecosystems (**Figure 4D**). This may be the result of simple geochemically available achiral molecules playing different roles, and in diverse metabolisms. This provides evidence of a more prominent role for achiral molecules in the largest organizational scale of metabolisms, found in communities of organisms. The critical transition between ecosystem-scale and individual-scale systems occurs when the number and distribution of achiral and chiral molecules are nearly equivalent.

A critical question is how the evolution of metabolism led to these distinctive patterns. Homochirality emerged via selection of sets of amino and nucleic acids with shared chiral orientation[13]. The most parsimonious explanation is that selection for members of these sets do not represent independent evolutionary events, but major transitions in molecular order governed by shared mechanisms of symmetry breaking within the reaction networks of nascent life[2,31]. For example, glutamate provides the amine source for 16 of the 20 coded amino acids, including glycine, which is achiral. The use of the same binding site by donor and acceptor in transamination preserves this chiral center in the synthesis of the 15 transaminated ketoacids that are chiral[2]. The chiral sugar, glyceraldehyde, plays a key role in transferring its chirality through several prebiotically plausible reactions[31], and these can even be influenced by amino acid catalysts[32]. These examples may represent specific instances of broader principles governing a large-scale role for propagation of broken chiral symmetries in the organization of biochemical reaction networks and metabolism.

To test this hypothesis, we classified reactions in our data by the number of chiral centers in both reactants and products, identifying the reaction chemistry as modulating or conserving the chiral complexity of molecules. We define three relevant reaction categories (SI Section 5): 1) **chiral modulating** reactions, where the reactants and products exhibit a net difference in the cumulative number of chiral centers (chiral complexity increases or decreases); 2) **chiral conserving** reactions, where the reactants and products exhibit an equal number of cumulative chiral centers (complexity is conserved) and 3) **achiral conserving** reactions, where there are no chiral centers in reactants nor in products (there is no chiral complexity). To explore as large a space of reactions as possible we used all reactions in CBR-db[61]. We find that a vast majority, 91,853 (75.96%), of metabolic reactions are chiral-conserving, while 22526 (18.63%) are chiral-modulating, and 6547 (5.41%) are achiral-conserving, as shown in **Figure 5A**. Chiral-conserving reactions appear very soon in the expansion process, dominating newly added reactions early-on and continuing to do so throughout the expansion (**Figure 5B**). This is consistent with the hypothesis of chiral transfer (where chiral centers are preserved along reaction pathways) playing an important role in the emergence of life[31,32], and it further corroborates how metabolic expansion led to increased selectivity of chiral-conserving reactions. Our results demonstrate, through large-scale statistics, how, rather than breaking reaction symmetries to generate new chiral centers, metabolism relies on conserving already broken molecular symmetries, propagating chiral centers along reaction pathways as a broad organizational principle of metabolic function and specificity.

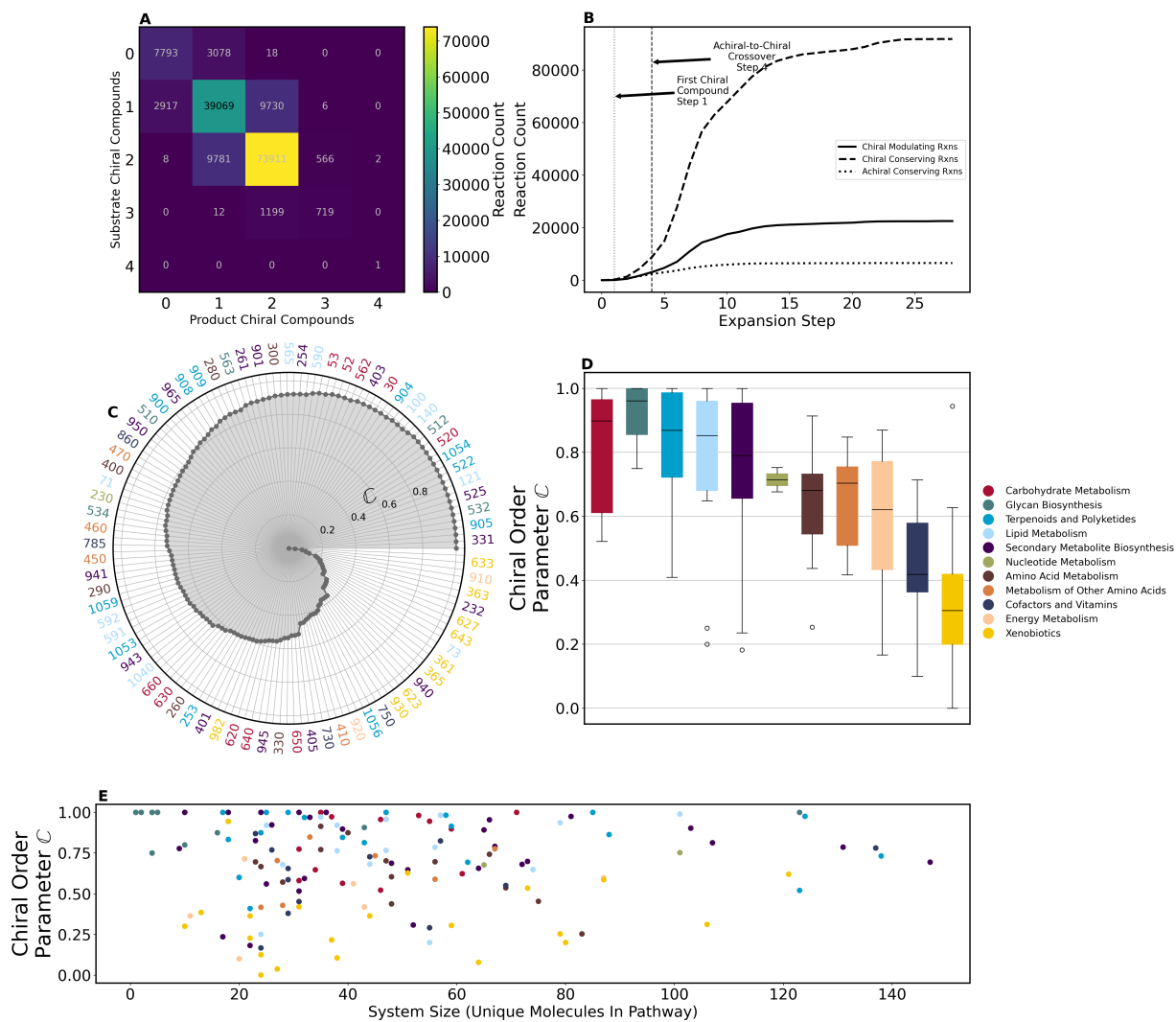

**Figure 5:** Metabolism conserves chiral centers, suggesting the propagation of broken chiral symmetries within reaction pathways. Only in aggregate do these pathways yield characteristics of the "chiral phase" of chemical network organization identified here as characteristic of functional metabolism. **A)** Frequency of biochemical reactions partitioned by the number of chiral reactants and products in each reaction. Heatmap values correspond to the count of all reactions in the CBR-db, representing the number of chiral reactants (y-axis) and chiral products (x-axis). **B)** For the primordial network expansion, cumulative counts of the three reaction chirality categories of chiral modulating, chiral conserving, and achiral conserving (defined in the text) show that chiral conserving reactions appear early and dominate the network expansion. **C)** The chiral order parameter, $\mathbb{C}$, varies significantly when metabolism is broken down into functional pathways, with values for individual pathways spanning the full range of allowable values for the chiral order parameter $0 \leq \mathbb{C} \leq 1$ (see also SI Figure S21). **D)** Grouping pathways into their higher-level functional modules confirms that subsets of metabolism associated with specific functionalities do not recapitulate the bulk chiral order parameter observed for organismal or ecosystem-scale metabolisms, with ranges of values outside those observed for individual or ecosystem-scale metabolic networks (compare to **Figure 4A**). Extreme outliers that fall far outside of whole-system (genome or metagenome scale network) values include pathways in Terpenoids and Polyketides,

known to be rich in complex chiral architectures. **E)** The chiral order parameter $\mathbb{C}$, when plotted for individual pathways by size (number of compounds in the pathway) does not exhibit scaling behavior, further distinguishing the scaling of $\mathbb{C}$ in biochemical networks as an emergent macroscale feature of a chiral phase of large-scale biochemical network organization.

Further insights are gained by looking at reaction pathways and functional modules within metabolism. Partitioning metabolic data by KEGG pathway[72] shows a widespread range in the chiral order parameter $\mathbb{C}$ across different functional modules within metabolism, from $\mathbb{C} = 0$ for atrazine degradation to $\mathbb{C} = 1$ for 18 pathways across metabolic classes, **Figure 5C**, revealing a partitioning of the conservation of chiral complexity (or lack thereof) along specific reaction pathways. Grouping pathways into their higher-level functional modules shows how subsets of metabolism associated with specific functions do not individually recapitulate the bulk chiral composition of a functioning organism or an ecosystem-scale metabolic community (**Figure 5D**). Nor do they recapitulate chiral scaling laws, perhaps due to their small sizes (**Figure 5E**, compared to **Figure 6A**). Modules and pathways also do not recapitulate bulk $\mathbb{C}$ of all known molecules, but seem to represent an intermediate scale of organization, where most metabolic pathways are constructed to conserve specific chiral/achiral molecular ratios as part of larger, functioning metabolisms. This indicates chiral scaling in organismal and ecosystem-scale metabolisms is indeed an emergent property of their functional organization as self-sustaining, living chemical networks.

## Discussion

Macromolecular homochirality is rightly a central focus of much origin of life research, but molecular chirality also plays a much broader role in the functioning of living systems[73–75]. In modern systems, molecules with chiral centers and chiral selective processes allow organisms to discriminate between mirror-image molecules and maintain precise control over metabolic processes[76,77]. Because of the fundamental role of chiral molecules in biology, chirality has even become an important branch of synthetic chemistry, shaping advancement in areas from materials science[78] to drug design[65–67,79]. Our study of the statistical patterns of chiral molecules in extant metabolisms provides hints of a potentially more extensive role of chiral selectivity as a general principle in the emergence and evolution of metabolism as well, where the structural distribution of chiral and achiral molecules could have played a critical role since the earliest organization of metabolic networks.

Many details to validate this hypothesis remain to be worked out. Our analyses offer a first attempt to integrate current understanding of the necessity of chiral selectivity emerging before LUCA with other pieces of the origin of life puzzle[7,8], such as how early metabolic networks could have evolved to break chiral symmetry leading to eventual selection of the macromolecules of life. Thus, our work provides a concrete step toward unifying metabolic and genetics-first approaches in origin of life research. Our focus on more universal chemical properties could, in turn, enable more precise timing of the major transitions in molecular organization, bridging geochemistry and LUCA, associated with chiral selectivity.

Our results are sufficiently general such that they can inform more universal models of the emergence of life from planetary geochemistry, than what is specific to the metabolisms or genetics of life as we know it. The expansion of evolving reaction networks, particularly into deeper regions of chemical space with larger molecules (more heavy atoms per molecule), requires selection on reaction pathways to make specific molecular configurations against an exponentially expanding space of potential molecular configurations[80]. Even the structural elucidation of chemical space, e.g., as in GDB, requires imposing selective constraints on bonds, elements, or molecular sizes of interest to make exploration of the space computationally tractable[81]. Any real-world chemistry must likewise impose selective constraints to generate highly complex molecules. We have shown how the transition from achiral to chiral-dominated chemistries can be elucidated by detailing how these molecules are embedded in a vast chemical space, and how reactions networks construct subsets of this space as they evolve. Features of this contingency in chemical space are recorded in the very molecular structures that compose these networks (here in terms of chirality). Understanding selectivity out of a vast unconstrained space of potential molecules is critical to understanding the network transitions that mediated the evolution from combinatorially explosive small-molecule geochemistry to LUCA, and should inform general principles for the emergence of life from any planetary environment.

Prior work suggests LUCA may have first emerged as ecosystem-scale, collectively-evolving reaction networks branched into individual lineages subject to Darwinian descent with modification (i.e. when functional, homochiral macromolecules became solidified in the networks)[1,82]. The hypothesis of an ecosystem-to-individual level transition in the emergence of life is consistent with our reported convergence of individual and ecosystem chiral order parameters, marking a major shift in chiral scaling in a range near the achiral-to-chiral phase boundary point at $\mathbb{C} = 0.5$. That is, the systematic size-dependencies we observe are suggestive of a potential progression whereby large-scale, predominantly achiral, ecosystem-level networks exist as a phase of chemical organization in one scaling regime (favoring achiral synthesis), which may have nucleated the formation of abundant chiral molecules. A phase transition in reaction network organization could be mediated by this chiral enrichment and symmetry-breaking by known mechanisms[13], marked by a new scaling regime (favoring conservation of chirality) due to selectivity along specific reaction pathways, with evolution toward increasing specificity that could have continued late into the emergence of life[68]. In our data, the remnants of LUCA that persist within modern metabolism constitute a small network with a high chiral order parameter ($\mathbb{C} = 0.65$), found in a range consistent with what is observed for small Archaea and Bacteria. This provides a first potential connection to be developed between network transitions in chiral selectivity, described by chiral scaling laws parameterized by $\mathbb{C}$, and the transition at the Darwinian threshold when individual lineages first solidified[1].

We find metabolism is enriched in large achiral compounds, relative to a random sampling of chemical space. Previous studies have suggested using enantiomeric excess as a biosignature[83], but our results indicate that searching for high mass achiral compounds could also be promising, as long as these are not oligomers. These heavy, achiral molecules are incredibly specific: for molecules with more than ~9 heavy atoms, adding, removing, or changing any bond is statistically likely to introduce a chiral center[40]. This suggests that heavy achiral compounds are the result of selection operating on the chemical systems in which they are found. Indeed, many of these molecules are highly evolved toxins, such as Fenbutatin oxide. Thus, a surprising conclusion of our analyses is that heavy, achiral molecules (i.e., those that are neither oligomeric nor composed of simple, repeating units) may serve as biosignatures. Additionally, since the scaling of the chiral order parameter $\mathbb{C}$ is a ubiquitous and conserved feature of metabolism, it too, provides a novel, system-level biosignature for life. The mechanisms underlying the specific scaling behaviors that characterize the chiral phase of network organization should be a subject of future theoretical and experimental work.

Our results demonstrate a previously unidentified hallmark characteristic of molecular organization in Earth's biology in the form of large-scale organization of achiral and chiral small molecules across metabolisms. Of most interest are insights that might bridge metabolism-first and genetics-first approaches to origins of life, as well as recent work indicating that the emergence of life is marked by a transition to the selective synthesis of abundant, high complexity molecules as measured by molecular assembly[80,84]. Interestingly, experimentally determined molecular biosignatures were identified in the form of complex molecules with molecular assembly larger than approximately 15 for covalently bonded molecules [84]. This corresponds roughly to >20 heavy atoms[84], and therefore is far above the achiral-to-chiral transition within chemical space. This connection suggests chiral selectivity, and the network transitions we report herein may be prerequisites to make complex molecules in abundance, detectable by their molecular assembly (e.g., where selection must act such that reaction networks reduce the entropy associated with increasing the abundance of specific chiral products). If the results presented here are indicative of universal properties of life, we should expect these patterns to be deeply connected to other universalities of selection[80] marking major transitions in molecular organization during the origin of life. Likewise, researchers across diverse fields may leverage these connections to enhance metabolic engineering, accelerate the formation of new evolutionary life forms in the lab, novel chiral drugs, and new chiral materials, as well as new chirality-based biosignatures, by identifying the underlying mechanisms for the macroscale order presented herein and emulating them in experimental systems.

# Acknowledgements

This work was supported by NASA grant number 80NSSC21K1402 and by the generous support of Schmidt Sciences to SIW through a science polymath fellowship. We wish to thank Steven

Glaser and William Bains for their constructive input during the early stages of this work, Michael Lachmann for constructive conversations and feedback and Mesa Shumacher for lending her artistic talents for the development of Figure 1A.

# Author Contributions

The initial conceptualization, supervision and funding of this work was provided by SIW. JM and LC contributed to conceptualization. JM developed the initial methodology, did initial investigations and SIW and JM wrote the initial draft. LS and CMathis aided SIW in project administration. JM, KN, LS, and CMilsaps developed final methodology and developed and implemented software for the analyses. KN, LC, and CMilsaps performed validation of datasets, and results. All authors contributed to reviewing and editing the manuscript.